\documentstyle[aps,twocolumn,prl,epsf]{revtex}

\begin{document}

\title{Effects of dilute nonmagnetic impurities on the 
${\bf Q}=(\pi,\pi)$ spin-fluctuation spectrum \\
in YBa$_2$Cu$_3$O$_7$}

\author{ N.\ Bulut }

\address{
Department of Mathematics, Ko\c{c} University, 
Istinye, 80860 Istanbul, Turkey} 

\date{\today} 
\draft

\twocolumn[\hsize\textwidth\columnwidth\hsize\csname@twocolumnfalse\endcsname
\maketitle 

\begin{abstract}
The effects of nonmagnetic impurities on the
${\bf Q}=(\pi,\pi)$ spin-fluctuation spectral weight 
${\rm Im}\,\chi({\bf Q},\omega)$ are studied
within the framework of 
the two-dimensional Hubbard model
using the random phase approximation. 
In the first part of the paper, 
the effects of the nonmagnetic impurities on the 
magnetic susceptibility
of the noninteracting ($U=0$) system,
$\chi_0({\bf q},\omega)$, are calculated 
with the self-energy and the vertex corrections 
using various forms of the 
effective electron-impurity interaction.
Here, the range and the strength of the impurity 
potential are varied as well as the concentration of the impurities.
It is shown that the main effect of dilute impurities 
on $\chi_0({\bf Q},\omega)$ is to cause a weak smearing.
In the second part, 
${\rm Im}\,\chi({\bf Q},\omega)$ 
is calculated for the interacting system.
Here, the calculations are concentrated on the processes 
which involve the impurity scattering of the spin fluctuations 
with finite momentum transfers.
In order to make comparisons with the experimental data
on the frequency dependence of ${\rm Im}\,\chi({\bf Q},\omega)$
in Zn substituted YBa$_2$Cu$_3$O$_7$, 
results are given for various values of 
the model parameters.

\end{abstract}

]

\section{Introduction and the Model}

Zn substitution has been used as a probe 
of the electronic properties in the layered cuprates 
and has yielded important results.
Small amount of Zn impurities lead to a strong suppression of
$d$-wave pairing \cite{Xiao}. 
Resistivity measurements find that Zn impurities 
act as strong scatterers \cite{Chien}.
Local magnetic measurements on Zn substituted cuprates 
have also yielded valuable information \cite{Mahajan,Mendels}.
It has been found that within the presence of Zn impurities 
the uniform susceptibility of YBa$_2$Cu$_3$O$_{7-\delta}$ 
follows a Curie-like temperature dependence.

Inelastic neutron scattering experiments find that Zn impurities 
cause important changes in the ${\bf Q}\equiv (\pi,\pi)$
spin-fluctuation spectrum \cite{Sidis,Fong,Regnault}.
The experiments have been carried out for 
2\% \cite{Sidis} and 0.5\% \cite{Fong} Zn 
concentrations. 
It is found that even
0.5\% Zn substitution in YBa$_2$Cu$_3$O$_7$ induces a peak in 
${\rm Im}\,\chi({\bf Q},\omega)$ at $\omega=40$ meV in the normal
state \cite{Fong}.
The $\omega$ width of the peak is about 10 meV and its width 
in momentum space is resolution limited. 
This peak becomes observable below $250^{\circ}$K and its intensity 
increases as $T$ is lowered without a discontinuity in its slope
at $T_c=87^{\circ}$K.
These results are quite different than those on pure 
YBa$_2$Cu$_3$O$_7$,
where the peak is observed only in the superconducting state
\cite{Rossat-Mignod,Mook,Fong95}.
While in the case of 0.5\% Zn substitution, 
no spectral weight is observed 
below $\sim 35$ meV, in the 
2\% Zn substituted sample there is significant amount 
of spectral weight down to 5 meV.
In this case, there is also a broad peak at 35 meV.
The momentum width of this peak is 0.5\AA$^{-1}$, 
which is about twice that of the peak in the 0.5\% Zn 
substituted and the pure samples.
Hence, the two main features induced in 
${\rm Im}\,\chi({\bf Q},\omega)$ in the normal state are the peak 
and the low frequency spectral weight 
observed in the 2\% Zn substituted sample. 
Both of these features become observable already above $T_c$.
The calculations reported in this article were carried out 
in order to investigate the origin of these two effects 
on the frequency  dependence of ${\rm Im}\,\chi({\bf Q},\omega)$
in YBa$_2$Cu$_3$O$_7$ in the normal state.

Here, the results of diagrammatic calculations on the effects 
of nonmagnetic impurities on ${\rm Im}\,\chi({\bf Q},\omega)$ 
obtained using the framework of 
the two-dimensional Hubbard model will be given.
In the first part of the paper, the effects of various types of
effective electron-impurity interactions on the magnetic 
susceptibility of the noninteracting ($U=0$) system,
$\chi_0({\bf Q},\omega)$, will be calculated.
In this case,  
both the strength and the range of the impurity interaction, 
in addition to the concentration of impurities, will be varied.
It will be found that in all these cases the main effect of 
impurity scattering is to cause a weak smearing of the 
structure in $\chi_0({\bf Q},\omega)$.
Hence, at this level, 
if $\chi_0$ is used in an RPA expression, 
$\chi({\bf Q},\omega)=\chi_0({\bf Q},\omega)
/(1-U\chi_0({\bf Q},\omega))$, then one obtains 
a smearing of ${\rm Im}\,\chi({\bf Q},\omega)$
by the impurity scattering rather than an enhancement
as seen in the experiments.
Here, it will be also noted that, for 
2\% impurity concentration, scatterings 
from an extended impurity potential 
could induce spectral weight at low frequencies.

In the next part of the paper, the effects of the processes 
where the spin fluctuations are scattered by the impurities with finite 
momentum transfers are calculated.
This type of processes
will be called the ``umklapp'' processes, since they involve 
the scattering of the spin fluctuations by the impurity potential 
with finite momentum transfers
as in the case of the scattering of the 
spin fluctuations by a charge-density-wave field. 
In this part, 
the effects of the umklapp scatterings will be estimated 
by calculating the irreducible off-diagonal susceptibility 
$\chi_0({\bf q},{\bf q'},\omega)$, where ${\bf q}\neq{\bf q'}$,
in the lowest order in the strength of the impurity potential.
It will be seen that the important umklapp processes 
are the ones which involve the transfer of momentum $2{\bf k}_F$
to the spin fluctuations, and that 
they lead to a peak in ${\rm Im}\,\chi({\bf Q},\omega)$ 
at $\omega_0=2|\mu|$, where $\mu$ 
is the chemical potential.
The underlying reason for this is a kinematic 
constraint which prohibits 
the creation of a particle-hole pair 
with center of mass momentum ${\bf q}=(\pi,\pi)-2{\bf k}_F$ 
and energy $\omega>\omega_0$. 
This constraint causes 
a nearly singular structure in 
$\chi_0({\bf Q},{\bf q},\omega)$ at $\omega=\omega_0$,
which in turn leads to the peak in 
${\rm Im}\,\chi({\bf Q},\omega)$ at $\omega_0$.
This effect has been noted previously \cite{Bulut}.
Here, 
results will be given for various sets of the model parameters
in order to make comparisons with the experimental data.
While in the pure system at low temperatures, 
the ${\bf Q}=(\pi,\pi)$ spin fluctuations are 
gapped below $\omega_0$, through this type of scatterings 
the ${\bf Q}=(\pi,\pi)$ spin fluctuations get mixed 
with the other wave vector components which are not gapped.
This process can also lead to finite spectral weight for 
$\omega<\omega_0$.
Comparisons with the experimental data suggest that 
the umklapp processes play an important role in determining the 
${\bf Q}=(\pi,\pi)$ spin dynamics in Zn substituted 
YBa$_2$Cu$_3$O$_7$.

The starting point is
the two-dimensional single-band Hubbard model given by 
\begin{eqnarray}
\label{Hubbard}
H=-t\sum_{\langle ij\rangle ,\sigma} 
(c^{\dagger}_{i\sigma}c_{j\sigma}
+c^{\dagger}_{j\sigma}c_{i\sigma})
+ U \sum_i 
c^{\dagger}_{i\uparrow} c_{i\uparrow} 
c^{\dagger}_{i\downarrow} c_{i\downarrow} \nonumber \\
-\mu \sum_{i,\sigma}
c^{\dagger}_{i\sigma} c_{i\sigma},
\end{eqnarray}
which 
will be used to model the spin fluctuations of the pure system.
Here $c_{i\sigma}$ ($c^{\dagger}_{i\sigma}$)
annihilates (creates) an electron with spin $\sigma$
at site $i$,
$t$ is the near-neighbor hopping matrix element and 
$U$ is the onsite Coulomb repulsion.
For simplicity, the hopping $t$,
the lattice constant $a$ and $\hbar$ are set to 1.

Within the presence of an impurity, 
the magnetic susceptibility is defined by 
\begin{equation}
\chi({\bf q},{\bf q'},i\omega_m) = 
\int_0^{\beta} d\tau \, 
e^{i\omega_m \tau}
\langle 
m^{-}({\bf q},\tau)  m^{+}({\bf q'},0) 
\rangle,
\label{chi}
\end{equation}
where $m^{+}({\bf q})= N^{-1/2} \sum_{{\bf p}}
c^{\dagger}_{{\bf p}+{\bf q}\uparrow} 
c_{{\bf p}\downarrow}$,
$m^{-}({\bf q})=(m^+({\bf q}))^{\dagger}$, 
and $\omega_m=2m\pi T$. 
By letting $i\omega_m\rightarrow \omega + i\delta$,
one obtains $\chi({\bf q},{\bf q'},\omega)$.
If one assumes that 
the effective interaction between an impurity and 
the electrons can be approximated by a static potential,
then the RPA expression for the magnetic 
susceptibility becomes
\begin{eqnarray} 
\label{rpa}
\chi({\bf q},{\bf q}',\omega) && = 
\chi_0({\bf q},{\bf q}',\omega) \nonumber \\
&& + 
U \sum_{{\bf q}''}
\chi_0({\bf q},{\bf q}'',\omega)
\chi({\bf q}'',{\bf q}',\omega),
\end{eqnarray}
where
$\chi_0({\bf q},{\bf q'},\omega)$ is the irreducible 
susceptibility dressed with the 
impurity scatterings.
The off-diagonal ${\bf q}\neq {\bf q'}$ terms of 
$\chi_0({\bf q},{\bf q'},\omega)$ vanish for the pure system, 
but they are finite within the presence of impurity scattering.
After obtaining $\chi({\bf q},{\bf q'},\omega)$ 
from Eq.~(\ref{rpa}), 
the averaging over the impurity location can be done, 
which sets 
${\bf q}={\bf q'}$.
$\chi({\bf q},{\bf q},\omega)$ obtained this way 
neglects the interactions between the impurities,
and here it will be assumed that these 
can be ignored in the dilute limit. 
In the following, for simplicity, 
$\chi({\bf q},{\bf q},\omega)$ 
and $\chi_0({\bf q},{\bf q},\omega)$ 
will be denoted by 
$\chi({\bf q},\omega)$ and $\chi_0({\bf q},\omega)$,
respectively.

In solving Eq.~(\ref{rpa}), if the 
off-diagonal terms are omitted, then one obtains
\begin{equation}
\label{rpa2}
\chi({\bf q},\omega) = 
{ \chi_0({\bf q},\omega) \over 
1 - U \chi_0({\bf q},\omega) },
\end{equation}
where $\chi_0({\bf q},\omega)$ is
dressed with the impurity scatterings.
In the next section, 
$\chi_0({\bf Q}=(\pi,\pi),\omega)$ 
will be calculated 
using various impurity potentials.
In the third section, 
$\chi({\bf Q},\omega)$ will be calculated without 
omitting the off-diagonal terms in Eq.~(\ref{rpa}).

\section{Effects of the nonmagnetic impurities without 
the umklapp scatterings}

In this part, the effects of dilute nonmagnetic impurities on 
the magnetic susceptibility of the noninteracting system
will be calculated 
using various impurity potentials.
The method used here for calculating $\chi_0({\bf Q},\omega)$ 
is similar to those used 
in Refs.~\cite{Hirschfeld,Quinlan,Li}.
Both the self-energy and the vertex corrections induced by the 
impurity scattering will be included \cite{Langer,Mahan}.

In Refs.~\cite{Poilblanc,Ziegler}, 
it has been shown how the electronic 
correlations lead to an extended effective interaction 
between an impurity and the electrons.
The extended nature of the effective impurity potential 
has been also emphasized in Ref.~\cite{Xiang}.
Assuming that it can be approximated by a
static form, the potential due to an impurity 
at site ${\bf r}_0$ can be written as 
\begin{equation}
\label{Veff}
V_{eff}=\sum_{\nu\alpha\sigma} V_{\nu\alpha} 
c^{\dagger}_{\nu\alpha\sigma} 
c_{\nu\alpha\sigma},
\end{equation}
where $\nu$ denotes the distance from the impurity and $\alpha$
denotes the different partial wave components.
The single-particle operators 
$c_{\nu\alpha\sigma}$ are given by 
\begin{equation}
c_{\nu\alpha\sigma} = 
\sum_{{\bf d}_{\nu}} 
g_{\nu\alpha}({\bf d}_{\nu}) 
c_{\sigma}({\bf r}_0+{\bf d}_{\nu}),
\end{equation}
where ${\bf d}_{\nu}$ sums over the sites at a distance 
$\nu$ away from the impurity 
and $g_{\nu\alpha}({\bf d}_{\nu})$'s 
are the coefficients of the partial-wave components.
In the following, an impurity interaction with a range
of $\sqrt{2}$ lattice spacings, which 
includes the second near-neighbor site,
will be considered.
For simplicity, subscript $i$ will be used 
to denote both $\nu$ and $\alpha$.

Within the presence of impurity scattering,  
the single-particle Green's function defined by 
\begin{equation}
G({\bf p},\tau) = - \langle T_{\tau} 
c_{{\bf p}\sigma} (\tau) 
c^{\dagger}_{{\bf p}\sigma} (0) \rangle
\end{equation}
is obtained from 
\begin{eqnarray}
\label{G}
G({\bf p},i\omega_n) && = G_0({\bf p},i\omega_n) \nonumber \\
&& + 
n_i
(G_0({\bf p},i\omega_n))^2
\sum_{ij} g_i({\bf p}) 
T_{ij}(i\omega_n) g_j({\bf p}),
\end{eqnarray}
where $n_i$ is the concentration of the impurities and 
the Matsubara frequency $\omega_n=(2n+1)\pi T$.
For an impurity potential with a range of $\sqrt{2}$ lattice spacings,
the subscripts $i$ and $j$ vary from 1 to 9.
The single-particle Green's function of the pure system
$G_0({\bf p},i\omega_n)$ entering Eq.~(\ref{G}) is given by
\begin{equation}
G_0({\bf p},i\omega_n)=
{1 \over 
i\omega_n-\varepsilon_{\bf p} }
\end{equation}
where the single-particle 
dispersion relation is
\begin{equation}
\varepsilon_{\bf p} = -2t (\cos{p_x} + \cos{p_y}) - \mu.
\end{equation}
The terms contributing to 
$G({\bf p},i\omega_n)$ are illustrated diagrammatically
in Fig.~1(a).
In Eq.~(\ref{G}),
$T_{ij}(i\omega_n)$ is the impurity-scattering $t$-matrix
and $g_i({\bf p})$'s are the form factors which are given by 
\begin{equation}
g_i({\bf p}) = \sum_{{\bf d}_{\nu}}
g_{\nu\alpha}({\bf d}_{\nu}) 
e^{i{\bf p}\cdot {\bf d}_{\nu}},
\end{equation}
where $i$ denotes $(\nu,\alpha)$.
The first five of the nine form factors 
used here are 
\begin{eqnarray}
g_{0s}({\bf p}) && = 1 \nonumber \\
g_{1s}({\bf p}) && = \cos{p_x}+\cos{p_y} \nonumber \\
g_{1p_x}({\bf p}) && = \sin{p_x} \nonumber \\
g_{1p_y}({\bf p}) && = \sin{p_y} \nonumber \\
g_{1d}({\bf p}) && = \cos{p_x}-\cos{p_y}
\end{eqnarray}
with similar expressions for the remaining
$\alpha=\sqrt{2}$ components having $s$, $p_x$, $p_y$ and 
$d$-wave symmetries.
The $t$-matrix $T_{ij}(i\omega_n)$ is obtained by solving 
\begin{equation}
T_{ij}(i\omega_n) = 
\delta_{ij}V_i + 
V_i \sum_{\ell} F_{i\ell}(i\omega_n) T_{\ell j}(i\omega_n),
\end{equation}
where
\begin{equation}
F_{ij}(i\omega_n) = {1\over N} \sum_{\bf p} 
g_i({\bf p}) G_0({\bf p},i\omega_n) g_j({\bf p}).
\end{equation} 

In order to calculate $\chi_0({\bf Q},i\omega_m)$, the irreducible 
interaction $\Gamma^0$ in the particle-hole channel 
due to impurity scattering is needed.
In Fig.~1(b), $\Gamma^0$ is illustrated diagrammatically 
and the corresponding expression is 
\begin{eqnarray}
\Gamma^0_{ii',jj'}&&({\bf p},{\bf p'},i\omega_n,i\omega_m) = 
- n_i g_i({\bf p}) g_{i'}({\bf p}+{\bf Q}) \nonumber \\
&& \times
T_{ij}(i\omega_n) T_{i'j'}(i\omega_n+i\omega_m) 
g_j({\bf p'}) g_{j'}({\bf p'}+{\bf Q}).
\end{eqnarray}
\begin{figure} 
\begin{center}
\leavevmode
\epsfxsize=7.5cm 
\epsfysize=7cm 
\epsffile{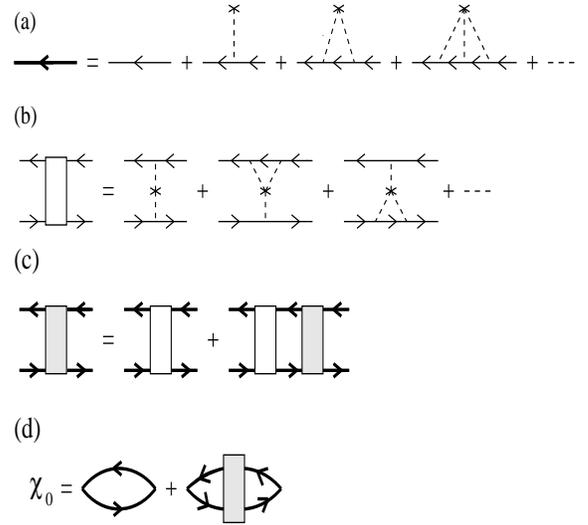}
\end{center}
\caption{
(a) Feynman diagrams for the single-particle Green's function 
dressed with the impurity scatterings.
(b) Irreducible vertex $\Gamma^0$ 
in the particle-hole channel arising from 
the impurity scattering.
(c) Reducible vertex $\Gamma$ in the particle-hole channel
given in terms of $\Gamma^0$.
(d) Irreducible magnetic susceptibility $\chi_0({\bf Q},i\omega_m)$ 
dressed with the impurity-induced self-energy and vertex corrections.
}
\end{figure}
\noindent 
Here, $\Gamma^0$ is evaluated with 
the center-of-mass momentum ${\bf Q}\equiv (\pi,\pi)$. 
Next,
the reducible vertex in the particle-hole channel 
is obtained by solving 
\begin{eqnarray}
\Gamma&&_{ii',jj'}({\bf p},{\bf p'},i\omega_n,i\omega_m) = 
\Gamma^0_{ii',jj'}({\bf p},{\bf p'},i\omega_n,i\omega_m) \nonumber \\
&& - \sum_{\ell\ell, ss'} 
{1\over N} \sum_{\bf k} 
\Gamma^0_{ii',\ell\ell'}({\bf p},{\bf k},i\omega_n,i\omega_m) 
G({\bf k},i\omega_n) \nonumber \\
&&\times
G({\bf k}+{\bf Q},i\omega_n+i\omega_m) 
\Gamma_{ss',jj'}({\bf k},{\bf p'},i\omega_n,i\omega_m),
\end{eqnarray}
which is illustrated in Fig.~1(c).
Here, it is noted that $\Gamma^0$ is calculated using the bare 
single-particle Green's function $G_0$, 
while $\Gamma$ is calculated using the 
Green's function dressed with the impurity scatterings.
This is necessary in order to prevent double-counting \cite{Mahan}.
In terms of the reducible vertex,
$\chi_0({\bf Q},i\omega_m)$ is given by 
\begin{eqnarray}
\label{chi0}
\chi_0({\bf Q},i\omega_m) =
\overline{\chi}&& _0({\bf Q},i\omega_m) \nonumber \\
- 
T\sum_{i\omega_n} {1\over N}\sum_{\bf p} 
{1\over N} &&\sum_{\bf p'} 
G({\bf p}+{\bf Q},i\omega_n+i\omega_m) G({\bf p},i\omega_n)
\nonumber \\ 
\times
\sum_{ii',jj'} 
\Gamma_{ii',jj'} &&({\bf p},{\bf p'},i\omega_n,i\omega_m) 
G({\bf p'}+{\bf Q},i\omega_n+i\omega_m) \nonumber \\
&&\times
G({\bf p'},i\omega_n).
\end{eqnarray}
Here, $\overline{\chi}_0({\bf Q},i\omega_m)$ 
includes only the impurity induced self-energy corrections, 
and it is given by 
\begin{eqnarray}
\overline{\chi}_0({\bf Q},i\omega_m) = - 
{T\over N} \sum_{{\bf p},i\omega_n} 
G({\bf p}+{\bf Q},i\omega_n+i\omega_m) \nonumber \\
\times
G({\bf p},i\omega_n).
\end{eqnarray}
These results in terms of the Matsubara frequencies
are analytically continued to the real frequency axis by the  
Pade approximation.
In the following, the results on $\chi_0({\bf Q},\omega)$ 
obtained this way will be compared with the 
Lindhard susceptibility of the pure system,
\begin{equation}
\label{Lindhard}
\chi^L_0({\bf Q},\omega) = 
{1\over N}
\sum_{\bf p}
{
f(\varepsilon_{{\bf p}+{\bf Q}}) -
f(\varepsilon_{\bf p}) \over 
\omega - ( \varepsilon_{{\bf p}+{\bf Q}} 
- \varepsilon_{\bf p} ) + i\delta}.
\end{equation}

Figure~2 shows results obtained using a strongly attractive
onsite impurity potential $V_0=-20$
and an impurity concentration of $n_i=0.02$.
In addition, here filling $\langle n\rangle=0.875$
and temperature $T=0.02$ are used.
In Fig.~2(a), $\chi_0({\bf Q},i\omega_m)$ versus $\omega_m$
is shown.
Also shown in this figure are 
$\overline{\chi}_0({\bf Q},i\omega_m)$, which does not include 
the impurity vertex corrections, and the Lindhard susceptibility 
$\chi_0^L({\bf Q},i\omega_m)$ of the pure system.
Figures~2(b) and (c) show the corresponding real 
and imaginary parts
obtained by the Pade analytic continuation.
Here, one observes that the impurity-induced 
self-energy corrections suppress 
$\chi_0({\bf Q},i\omega_m)$ and 
smear the structure in $\chi_0({\bf Q},\omega)$.
For instance, the hump in 
${\rm Re}\,\chi_0^L({\bf Q},\omega)$ at 
$\omega\approx 2|\mu|=0.48$ is smeared 
by the self-energy corrections, but when 
the vertex corrections are included, the effect 
of the 
\begin{figure} 
\begin{center}
\leavevmode
\epsfxsize=9cm 
\epsfysize=10cm 
\epsffile{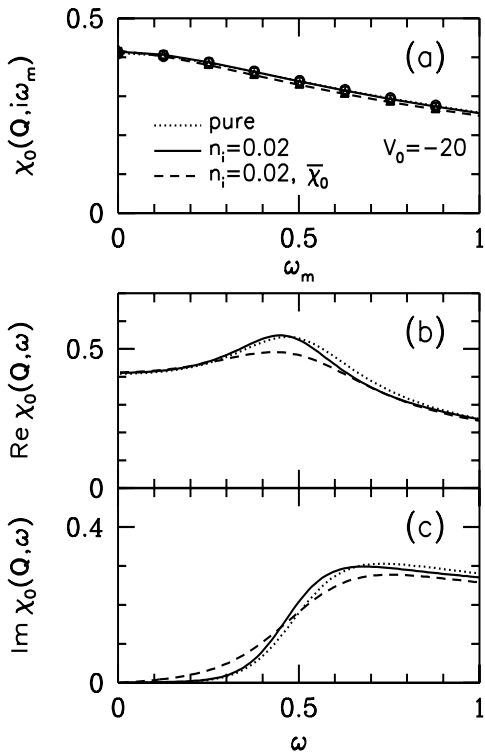}
\end{center}
\caption{
Results on (a) $\chi_0({\bf Q},i\omega_m)$ versus $\omega_m$,
(b) ${\rm Re}\,\chi_0({\bf Q},\omega)$ versus $\omega$ and 
(c) ${\rm Im}\,\chi_0({\bf Q},\omega)$ versus $\omega$
for an onsite impurity potential with $V_0=-20$ and 
impurity concentration $n_i=0.02$.
These results were obtained for 
$\langle n\rangle =0.875$ and $T=0.02$.
}
\end{figure}
\noindent    
self-energy corrections is nearly canceled.
This hump is because of a logarithmic singularity in 
${\rm Re}\,\chi_0({\bf Q},\omega)$ at $T=0$
originating from the dynamic nesting of the Fermi surface.
In a real system, 
the deviations from the simple tight-binding model 
which has 
only near-neighbor hoppings would lead to a 
suppression of this hump.
In addition, 
the scattering of the 
single-particle excitations by the spin-fluctuations 
would suppress it also.
In Figs.~2(b) and (c), the difference between the solid and 
the dotted lines is of order the resolution of the 
Pade analytic continuation.
These calculations were repeated using 
$n_i=0.005$ instead of $n_i=0.02$,
in which case the 
difference between 
$\chi_0$ and $\chi_0^L$ 
becomes even smaller (not shown here).

Figure~3 shows results also for $n_i=0.02$
and an onsite impurity potential but now with $V_0=-1$.
One observes that in this case the impurities 
have a stronger effect on $\chi_0$.
In addition, it is noted that 
${\rm Re}\,\chi_0({\bf Q},\omega)$ for 
$\omega\sim 0.5$ is suppressed,
and ${\rm Re}\,\chi_0({\bf Q},\omega\sim 0)$ 
gets enhanced by a small amount.
Among the various forms tried for the impurity potential,
this is the only case where an enhancement of 
$\chi_0({\bf Q},\omega)$ by the impurity scatterings has been obtained. 
This effect is due to the enhancement of the single-particle 
density of states at the impurity site by the attractive potential.
The enhancement of 
${\rm Re}\,\chi_0({\bf Q},\omega\sim 0)$ at small frequencies 
could lead to some enhancement of the low-frequency 
antiferromagnetic spin 
\begin{figure} 
\begin{center}
\leavevmode
\epsfxsize=9cm 
\epsfysize=10cm 
\epsffile{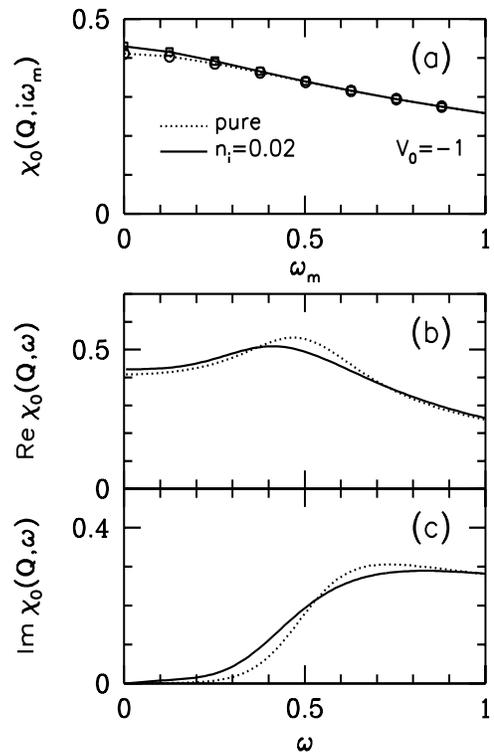}
\end{center}
\caption{
Results similar to those in Fig.~2 but for $V_0=-1$.
}
\end{figure}
\noindent
fluctuations for a system with a 
large Stoner factor.
However, it is seen that 
${\rm Re}\,\chi_0({\bf Q},\omega)$ 
for $\omega$ near $\omega_0=2|\mu|$ 
is suppressed, and actually it would not be possible 
to explain the peak in the neutron scattering data with these results. 
This calculation was repeated using $V_0=1$, and a suppresion 
of ${\rm Re}\,\chi_0({\bf Q},\omega)$ was obtained 
due to the depletion of the 
single-particle density of states
(not shown here).
    
Figure 4 shows results on $\chi_0$ obtained 
using an extended impurity potential 
for $n_i=0.02$ and 0.005.
These results were obtained for a
potential with a 
range of $\sqrt{2}$ lattice spacings and
with the following parameters:
$V_{0s}=-20$, $V_{1\alpha}=0.5$ and 
$V_{\sqrt{2}\alpha}=-0.25$ where
$\alpha$ denotes the $s$, $p_x$, $p_y$ and $d$-wave components.
These values for $V_{\nu\alpha}$'s are comparable to those obtained in 
Ref.~\cite{Ziegler}.
The calculations were repeated using various 
other values for the $V_{\nu\alpha}$'s, and
it has been found that small changes in $V_{\nu\alpha}$'s do not 
change the results shown here significantly.
For instance, increasing $V_{\nu\alpha}$'s by 50\% does
not change the conclusions of this section. 
In Fig.~4,
it is seen that for $n_i=0.02$ an extended potential leads to 
significant smearing of the structure in 
$\chi_0({\bf Q},\omega)$.
In this case, 
the hump in ${\rm Re}\,\chi_0({\bf Q},\omega)$ is 
rounded off, and spectral weight is induced for 
$\omega < \omega_0$.
Hence, comparing 
with Fig.~2, one observes that while an onsite impurity potential 
\begin{figure} 
\begin{center}
\leavevmode
\epsfxsize=9cm 
\epsfysize=10cm 
\epsffile{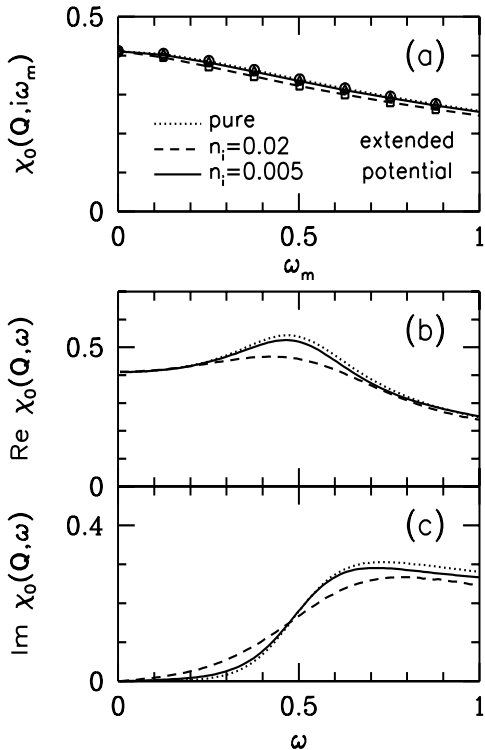}
\end{center}
\caption{
Results similar to those in Fig.~2 but for an extended 
potential with the following set of parameters:
$V_{0s}=-20$, $V_{1\alpha}=0.5$ and  $V_{\sqrt{2}\alpha}=-0.5$.
Here results are given for both $n_i=0.02$ and 0.005.
}
\end{figure}
\noindent
does not lead to spectral weight 
for $\omega < \omega_0$, an extended potential induces 
spectral weight in the gap.
Also shown in Fig.~4 are the results
for $n_i=0.005$, in which case the effect of the impurities on 
$\chi_0({\bf Q},\omega)$ is weaker.
The fact that 0.5\% impurities induce less  
spectral weight in the gap compared to the 2\% case 
is consistent with the neutron scattering data 
by Refs.~\cite{Sidis,Fong}. 
However, even for $n_i=0.02$, the amount of the 
spectral weight induced in the gap is small.
In the next section, it will be seen that 
the umklapp processes could also contribute to 
${\rm Im}\,\chi({\bf Q},\omega)$ for $\omega < \omega_0$.

From the results presented here,
one observes that if $\chi_0({\bf Q},\omega)$ computed 
in this section is used in Eq.~(\ref{rpa2}), 
which omits the umklapp scatterings, then one would only 
obtain a smearing of ${\rm Im}\,\chi({\bf Q},\omega)$
by the impurities.
Hence, it would not be possible to explain how
0.5\% Zn 
impurities induce a peak in 
${\rm Im}\,\chi({\bf Q},\omega)$ in the normal state
of YBa$_2$Cu$_3$O$_7$.
In the next section, the effects of 
the impurity scatterings 
with finite momentum transfers will be taken into account.

\section{Effects of the impurity induced umklapp scatterings}

Here, ${\rm Im}\,\chi({\bf Q},\omega)$ will be calculated 
without omitting the off-diagonal components 
in Eq.~(\ref{rpa}).
These off-diagonal terms will be calculated 
in the lowest order in the strength of the impurity potential, 
as illustrated diagrammatically in Fig.~5(a).
In the previous section, 
it was found that 0.5\% and 2\% impurities cause 
only a weak smearing of $\chi_0({\bf Q},\omega)$.
For this reason, the diagonal components of Eq.~({\ref{rpa})
will be approximated by the Lindhard susceptibility 
$\chi_0^L$.
This will not change the nearly singular contribution originating 
from the umklapp scatterings at $\omega\approx\omega_0$.
The expression for 
$\chi_0({\bf Q},{\bf q},i\omega_m)$ 
corresponding to the diagrams shown in Fig.~5(a) is  
\begin{eqnarray} 
\label{chi012}
\chi_0({\bf Q},{\bf q},i\omega_m) = 
-V_0 {T\over N} && \sum_{{\bf p},i\omega_n} 
\bigg\{ 
G_0({\bf p}+{\bf Q},i\omega_n+i\omega_m) \nonumber \\
\times 
G_0({\bf p},i\omega_n) 
&& G_0({\bf p}+{\bf Q}-{\bf q},i\omega_n) \nonumber \\
+ 
G_0({\bf p},i\omega_n) 
&& G_0({\bf p}+{\bf q},i\omega_n+i\omega_m) \nonumber \\
\times
&& G_0({\bf p}+{\bf Q},i\omega_n+i\omega_m)
\bigg\}.
\end{eqnarray}
Here ${\bf Q}\equiv (\pi,\pi)$ 
and ${\bf q}={\bf Q}-{\bf Q^*}$, where 
${\bf Q^*}$ is the momentum transferred during 
the scattering from the impurity.
Upon carrying out the summation over 
$i\omega_n$ in Eq.~(\ref{chi012})
and letting $i\omega_m\rightarrow \omega+i\delta$,
one obtains
\begin{eqnarray}
\label{chi012b}
\chi_0({\bf Q},{\bf q},\omega) && = 
A({\bf Q},{\bf q},\omega) 
+
A^*({\bf Q},{\bf q},-\omega) \nonumber \\
&& +
B({\bf Q},{\bf q},\omega) 
-
B^*({\bf Q},{\bf q},-\omega),
\end{eqnarray}
where ``*'' stands for complex conjugation, and 
$A$ and $B$ are given by 
\begin{eqnarray}
\label{A}
A({\bf Q},{\bf q},\omega) = 
- {V_{\bf k} \over N} 
\sum_{\bf p} 
\bigg\{ 
{ f(\varepsilon_{{\bf p}+{\bf Q^*}}) \over 
\omega - 
( \varepsilon_{{\bf p}+{\bf Q}} - 
\varepsilon_{{\bf p}+{\bf Q^*}} ) + i\delta } \nonumber \\
-
{ f(\varepsilon_{\bf p}) \over 
\omega - 
( \varepsilon_{{\bf p}+{\bf Q}} - 
\varepsilon_{\bf p} ) + i\delta } 
\bigg\} 
{1 \over 
\varepsilon_{{\bf p}+{\bf Q^*}} - \varepsilon_{\bf p} }
\end{eqnarray}
\begin{eqnarray}
\label{B}
&& B({\bf Q},{\bf q},\omega) = 
- {V_{\bf k} \over N} 
\sum_{\bf p} 
{1 \over 
\varepsilon_{{\bf p}+{\bf Q^*}} - \varepsilon_{\bf p} } \nonumber \\
&& \times
{ f(\varepsilon_{{\bf p}+{\bf Q^*}}) 
\over 
(\omega - 
( \varepsilon_{{\bf p}+{\bf Q}} - 
\varepsilon_{{\bf p}+{\bf Q^*}} ) + i\delta )
( \omega - 
( \varepsilon_{{\bf p}+{\bf Q}} - 
\varepsilon_{\bf p} ) + i\delta ) }. 
\end{eqnarray}

A general form for the effective interaction between 
the electrons and an impurity located at site ${\bf r}_0$
is given by
\begin{equation}
V_{eff} =
\label{V}
{1\over N}\sum_{{\bf k}}
e^{i{\bf k}\cdot {\bf r}_0}
V_{{\bf k}} \sum_{{\bf p},\sigma} 
c^{\dagger}_{{\bf p}+{\bf k}\sigma} c_{{\bf p}\sigma}.
\end{equation}
Below, it will be seen that 
for ${\bf Q^*}$ near $2{\bf k}_F$,
$\chi_0({\bf Q},{\bf q},\omega)$ has a nearly singular structure at 
$\omega\approx\omega_0$, while for  
${\bf Q^*}$ away from $2{\bf k}_F$,
$\chi_0({\bf Q},{\bf q},\omega)$
is a smooth function of $\omega$ with a small amplitude.
Hence,
$\chi({\bf Q},\omega)$ will be 
calculated using only the $\pm{\bf Q^*}$ components of 
the effective impurity interaction, 
\begin{equation}
\label{V2}
V_0\sum_{{\bf p}\sigma} 
(c^{\dagger}_{{\bf p}+{\bf Q^*}\sigma}
c_{{\bf p}\sigma}
+
c^{\dagger}_{{\bf p}-{\bf Q^*}\sigma}
c_{{\bf p}\sigma}),
\end{equation}
where $V_0=V_{\bf Q^*}$ is taken as a parameter.
This is necessary, since, in order to have sufficient 
frequency resolution, the calculation
needs to be carried out on a large lattice, 
which is difficult to do using directly 
Eq.~(\ref{V}).
Furthermore, 
the detailed ${\bf k}$ dependence of $V_{\bf k}$ 
is not known, especially for ${\bf k}$ near 
$2{\bf k}_F$.
The scattering of a quasiparticle with ${\bf Q^*}=2{\bf k}_F$ 
momentum transfer
is sketched in the Brillouin zone in Fig.~5(b).
Using the interaction given 
\begin{figure} 
\begin{center}
\leavevmode
\epsfxsize=6cm 
\epsfysize=2.5cm 
\epsffile{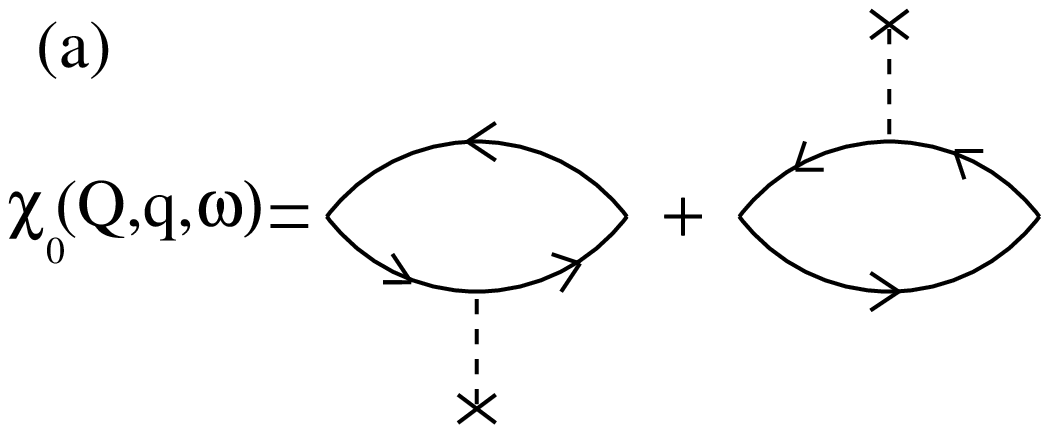}
\end{center}
\begin{center}
\leavevmode
\epsfxsize=6cm 
\epsfysize=4cm 
\epsffile[18 205 592 598]{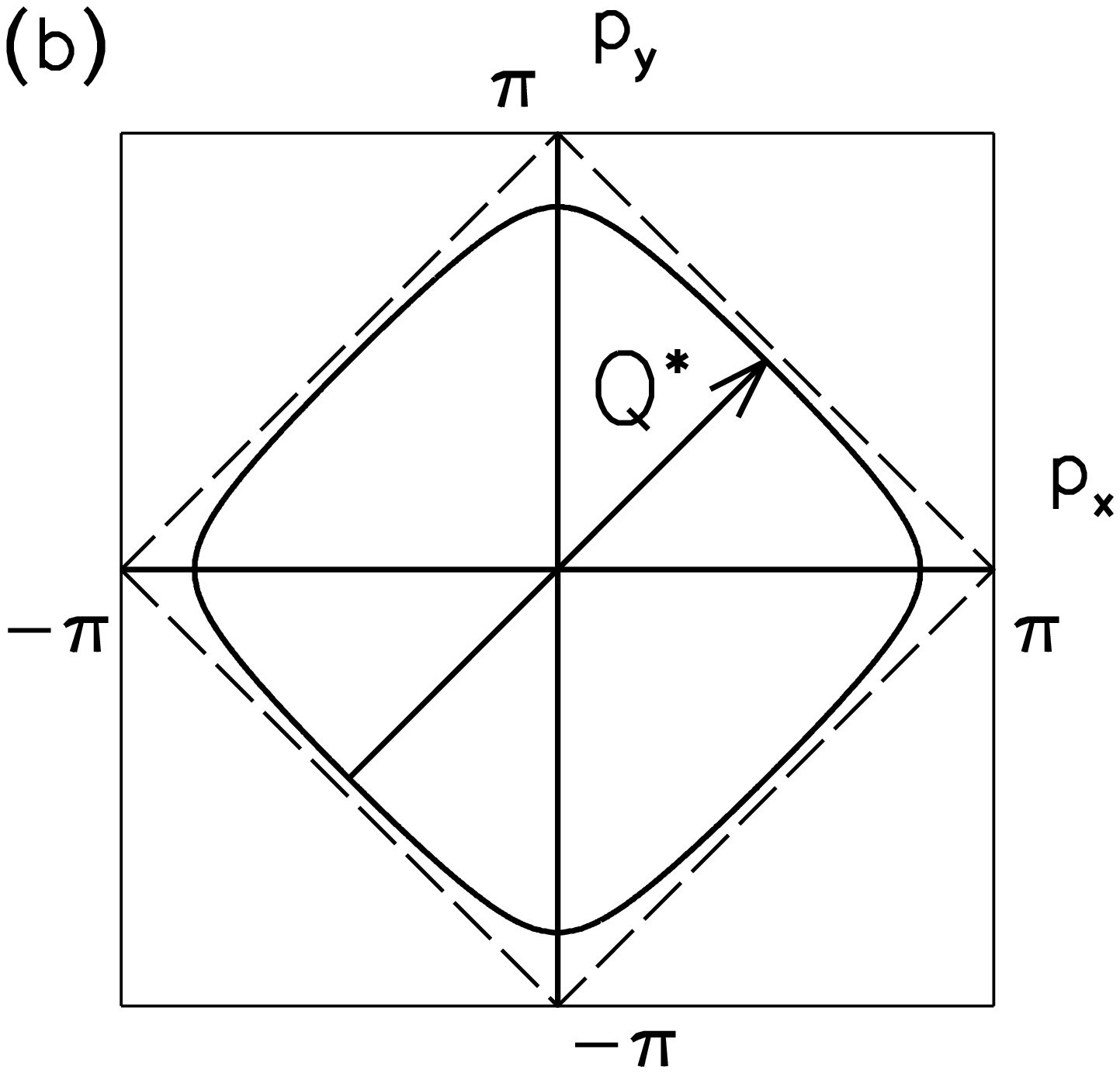}
\end{center}
\caption{
(a) Feynman diagrams for the 
lowest-order terms contributing to 
the off-diagonal irreducible susceptibility
$\chi_0({\bf Q},{\bf q},\omega)$.
(b) Sketch in the Brillouin zone of a quasiparticle 
scattering process where momentum 
$2{\bf k}_F$ is transferred.
}
\end{figure}
\noindent    
in Eq.~(\ref{V2}), 
one obtains for $\chi({\bf Q},\omega)$,
\begin{eqnarray} 
\label{chi22}
\chi({\bf Q}&&,\omega) = 
\Big\{\chi^L_0({\bf Q},\omega) 
(1 - U\chi^L_0({\bf q},\omega)) \nonumber \\
&&+ 
4 U(\chi_0({\bf Q},{\bf q},\omega))^2 \Big\} 
\times \Big\{
(1-U\chi^L_0({\bf Q},\omega)) \nonumber \\
&&\times
(1 - U\chi^L_0({\bf q},\omega)) 
- 4 (U\chi_0({\bf Q},{\bf q},\omega))^2\Big\}^{-1},
\end{eqnarray}
where 
${\bf q}={\bf Q}-{\bf Q^*}$.
Here,
the factor of 4 multiplying 
$(\chi_0({\bf Q},{\bf q},\omega))^2$ 
is to take into account the 
scatterings with momentum transfers 
$(\pm Q^*,\mp Q^*)$ in addition to 
$(\pm Q^*,\pm Q^*)$, where ${\bf Q^*}=(Q^*,Q^*)$. 

In the following, results will be shown for 
$\langle n\rangle =0.86$ and $T=0.05$,
in which case $\omega_0=2|\mu|\approx 0.55$.
Figure~6(a) shows the real and the imaginary parts of 
$\chi_0({\bf Q},{\bf q},\omega)$ for 
${\bf Q^*}=2{\bf k}_F$. 
Here, the Fermi wave vector ${\bf k}_F$ 
has been taken along 
(1,1) for simplicity.
In evaluating $\chi({\bf Q},\omega)$ 
with Eq.~(\ref{chi22}), the Lindhard susceptibilities 
$\chi_0^L({\bf Q},\omega)$ and 
$\chi_0^L({\bf q}={\bf Q}-{\bf Q^*},\omega)$ are also used
and, hence, they are plotted in Figures~6(b) and (c).
Here, one notes the similarity 
between the $\omega$ dependence of 
$\chi_0({\bf Q},{\bf q},\omega)$ 
and $\chi_0^L({\bf q}={\bf Q}-{\bf Q^*},\omega)$.
Both have vanishing spectral weight 
for $\omega > \omega_0$ because of
kinematic constraints. 

In Fig.~7, 
${\rm Im}\,\chi({\bf Q},\omega)$ versus $\omega$ obtained 
from Eq.~(\ref{chi22})
by using the results of Fig.~6 are 
shown for different values of $U$.
The solid lines are for $V_0=0.05$ and the dashed lines are for 
$V_0=0$ corresponding to the pure case.
Here, 
\begin{figure} 
\begin{center}
\leavevmode
\epsfxsize=9cm 
\epsfysize=10cm 
\epsffile{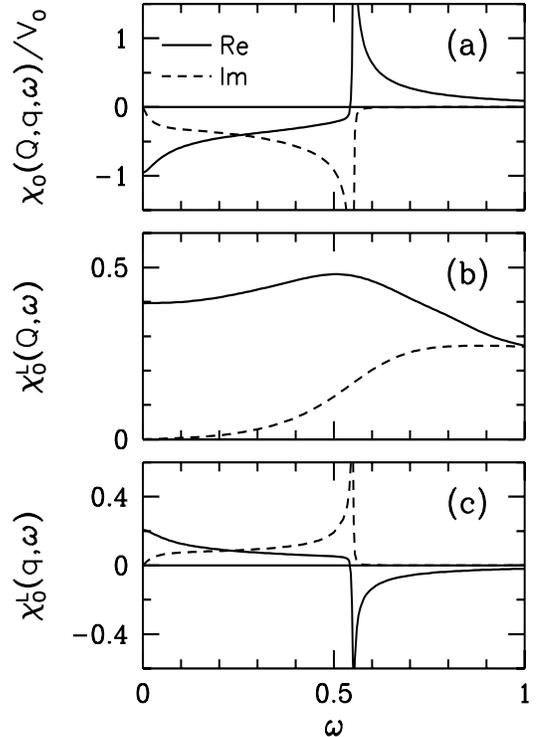}
\end{center}
\caption{
Frequency dependence of 
(a) $\chi_0({\bf Q},{\bf q},\omega)$
(b) $\chi^L_0({\bf Q},\omega)$ and 
(c) $\chi^L_0({\bf q},\omega)$.
Here ${\bf q}={\bf Q}-{\bf Q^*}$ and ${\bf Q^*}=2{\bf k}_F$.
These results were obtained for 
$\langle n\rangle =0.86$ and $T=0.05$.
}
\end{figure}
\noindent
\begin{figure} 
\begin{center}
\leavevmode
\epsfxsize=9cm 
\epsfysize=10cm 
\epsffile{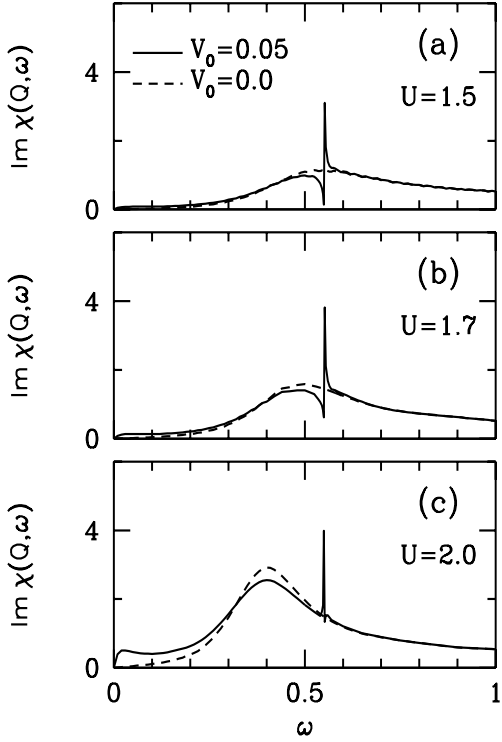}
\end{center}
\caption{
${\rm Im}\,\chi({\bf Q},\omega)$ versus $\omega$ for 
(a) $U=1.5$, (b) $U=1.7$, and (c) $U=2.0$.
These results were for $V_0=0.05$ and 0.
If the value of $V_0$ is increased to 0.06, the sharp peak at 
$\omega_0\approx 0.55$ diverges.
}
\end{figure}
\noindent
it is seen that a peak is induced at 
$\omega_0\approx 0.55$ by turning on $V_0$.
One also observes that as $U$ is increased, a hump 
develops below the peak.
This is because of the 
RPA enhancement of $\chi$
in the pure case.
In a real system, 
it is expected that this hump will be smaller because of
the band-structure effects and the damping of the 
quasiparticles by the spin fluctuations.
In Fig.~8, it is also seen that 
the impurity contribution to the 
low frequency part of ${\rm Im}\,\chi({\bf Q},\omega)$ 
increases as $U$ is increased.
In addition, 
in Fig.~6(c) it was seen that 
$\chi_0({\bf q}={\bf Q}-2{\bf k}_F,\omega)$ 
has sharp structure near $\omega_0$, 
but this is not responsible for producing the peak in 
${\rm Im}\,\chi({\bf Q},\omega)$. 
For instance, taking out the factor of 
$(1-U\chi_0^L({\bf q},\omega))$ from both the numerator 
and the denominator in Eq.~(\ref{chi22}) does not 
change the structure of ${\rm Im}\,\chi({\bf Q},\omega)$. 

Next, in order to have a better understanding 
of these results, 
a sketch of the important wave vectors and frequencies 
are given in the ${\bf q}$-$\omega$ plane
in Fig.~8. 
Here, ${\bf Q}=(\pi,\pi)$,
and ${\bf q}$ and $2{\bf k}_F$ are taken along (1,1).
The vertical dashed line denotes ${\bf q}={\bf Q}-2{\bf k}_F$ 
and the horizontal dashed line is for $\omega_0=2|\mu|$.
The shaded area represents the region where 
${\rm Im}\,\chi_0^L({\bf q},\omega)\neq 0$ at $T=0$.
Here, one observes that the ${\bf Q^*}=2{\bf k}_F$ scattering 
of the ${\bf Q}=(\pi,\pi)$ spin fluctuations will 
lead to a mixing with the ${\bf q}={\bf Q}-2{\bf k}_F$ 
component of the spin fluctuations, which has 
spectral weight only for $\omega < \omega_0$.
A general 
\begin{figure} 
\begin{center}
\leavevmode
\epsfxsize=6cm 
\epsfysize=6cm 
\epsffile{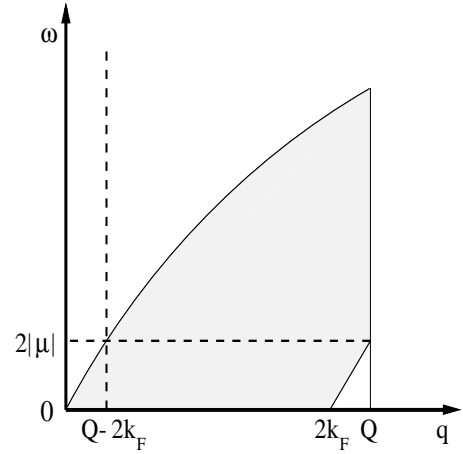}
\end{center}
\caption{
Sketch in ${\bf q}-\omega$ plane of 
the important wave vectors and frequencies (not to scale).
The vertical dashed line denotes 
${\bf q}={\bf Q}-2{\bf k}_F$ 
where ${\bf Q}=(\pi,\pi)$
and the horizontal dashed line denotes 
$\omega_0=2|\mu|$. 
The shaded area represents the region where 
${\rm Im}\,\chi^L_0({\bf q},\omega)\neq 0$
at $T=0$.
Here ${\bf q}$ and $2{\bf k}_F$ were taken along $(1,1)$.
}
\end{figure}
\noindent
impurity potential such as Eq.~(\ref{V})
would lead to a mixing with all wave vectors.
Hence, the umklapp processes could also 
contribute to ${\rm Im}\,\chi({\bf Q},\omega)$ for 
$\omega < \omega_0$, 
in addition to the results 
seen in the previous section.

In order to show that the nearly singular $\omega$ dependence of 
$\chi_0({\bf Q},{\bf q},\omega)$ occurs only for 
${\bf Q^*}\approx 2{\bf k}_F$, in Fig.~9 results 
on $\chi_0({\bf Q},{\bf q},\omega)/V_0$ 
are shown for various values of ${\bf Q^*}$.
As seen in Fig.~9(a),
for ${\bf Q^*}=1.02(2{\bf k}_F)$, 
there is a sharp structure in 
$\chi_0({\bf Q},{\bf q},\omega)$ at $\omega$ less 
than $\omega_0$, while for ${\bf Q^*}=0.98(2{\bf k}_F)$, 
this occurs at $\omega>\omega_0$.
As ${\bf Q^*}$ moves away from 
$2{\bf k}_F$, the position of the structure in 
$\chi_0({\bf Q},{\bf q},\omega)$ shifts away from $\omega_0$ and its 
amplitude decreases.
In Figs.~9(b) and (c), 
results on
${\rm Re}\,\chi_0({\bf Q},{\bf q},\omega)/V_0$ 
versus $\omega$ are shown for 
${\bf Q^*}$ along (1,1) and (1,0), respectively.
Here, ${\rm Im}\,\chi_0({\bf Q},{\bf q},\omega)/V_0$ is not shown 
since it is a smooth function of $\omega$ with amplitude less
than 0.05.
In these figures, it is seen that the magnitude of 
$\chi_0({\bf Q},{\bf q},\omega)$ at $\omega\approx \omega_0$
is considerably smaller when ${\bf Q^*}$ 
is away from $2{\bf k}_F$.
This 
supports the use of only the ${\bf Q^*}$  component
of $V_{eff}$ in solving for $\chi({\bf Q},\omega)$.
If a general potential such as Eq.~(\ref{V}) instead of 
Eq.~(\ref{V2}) were used in solving for 
$\chi({\bf Q},\omega)$, then the peak in 
${\rm Im}\,\chi({\bf Q},\omega)$ would again occur at $\omega_0$ 
but with a broadened width
because of the contributions originating from scatterings 
with ${\bf Q^*}$ slightly away from $2{\bf k}_F$.

In Fig.~7, 
one notes that the line shape of the peak depends 
on the value of $U$.
For $U=1.5$ and 1.7, the peak is asymmetric;
there is a dip below the peak.
For $U=2.0$, the dip is not observed.
In order to have a better 
understanding of this, further results 
on the line shape are shown 
in Figures~10(a) and (b) 
for $U=1.7$ and 2.0, respectively.
In these figures,
if the value of $V_0$ is increased to 0.06, 
the peak at $\omega_0$ diverges.
Also shown here in Fig.~10(c) is the quantity 
\begin{figure} 
\begin{center}
\leavevmode
\epsfxsize=9cm 
\epsfysize=10cm 
\epsffile{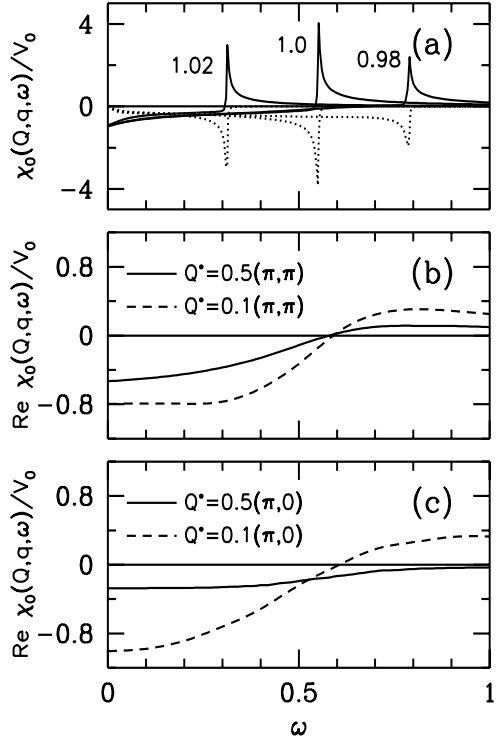}
\end{center}
\caption{
(a) $\chi_0({\bf Q},{\bf q},\omega)/V_0$ versus $\omega$ for 
different values of ${\bf Q^*}$ near $2{\bf k}_F$ along (1,1).
Here, the numbers next to the curves indicate the magnitude 
of ${\bf Q^*}$ in units of $2k_F$.
The solid and the dotted curves represent the real and 
the imaginary parts.
In (b) and (c), 
results on
${\rm Re}\,\chi_0({\bf Q},{\bf q},\omega)/V_0$ 
versus $\omega$ are plotted for ${\bf Q^*}$ away from $2{\bf k}_F$.
}
\end{figure}
\noindent
\begin{equation}
\gamma(\omega) = 4U^2 (\chi_0({\bf Q},{\bf q},\omega))^2,
\end{equation}
which enters Eq.~(\ref{chi22}).
For $U=1.7$, the system is away from 
the magnetic instability and, 
in this case, ${\rm Re}\,\gamma(\omega)$ 
has a stronger effect than the imaginary part.
While 
for $\omega > \omega_0$, 
${\rm Re}\,\gamma(\omega)$ leads to a stronger RPA 
enhancement of ${\rm Im}\,\chi({\bf Q},\omega)$, for 
$\omega < \omega_0$ it suppresses 
${\rm Im}\,\chi({\bf Q},\omega)$.
On the other hand, for $U=2$,
where $1-U{\rm Re}\,\chi_0^L({\bf Q},\omega)$ is small, 
the pure system is already close to the magnetic instability
and $U{\rm Im}\,\chi_0^L({\bf Q},\omega)$ 
acts as a damping of the RPA enhancement.
In this case, 
${\rm Im}\,\gamma(\omega)$ becomes more important
and it leads to the peak in 
${\rm Im}\,\chi({\bf Q},\omega)$ by suppressing 
the damping of the RPA enhancement.
Possibly, this is the case more applicable to 
YBa$_2$Cu$_3$O$_7$.

The results seen in Fig.~10
were obtained using a finite broadening 
$\delta=0.001$ in Eqs.~(\ref{Lindhard}),
(\ref{A}) and (\ref{B}).
For comparison, 
results obtained using $\delta=0.01$ are shown
in Fig.~11.
In this case, the structure in $\gamma(\omega)$ 
and the 
peak in ${\rm Im}\,\chi({\bf Q},\omega)$ 
are broader.
It is also seen that 
the integrated spectral weight 
in the peak increases with $\delta$ for $U=2$.
If $\delta$ is increased further, 
the width of the peak 
in ${\rm Im}\,\chi({\bf Q},\omega)$ 
for $U=2$ continues to increase (not shown here).
It is noted that the 
scattering of the quasiparticles
\begin{figure} 
\begin{center}
\leavevmode
\epsfxsize=9cm 
\epsfysize=10cm 
\epsffile{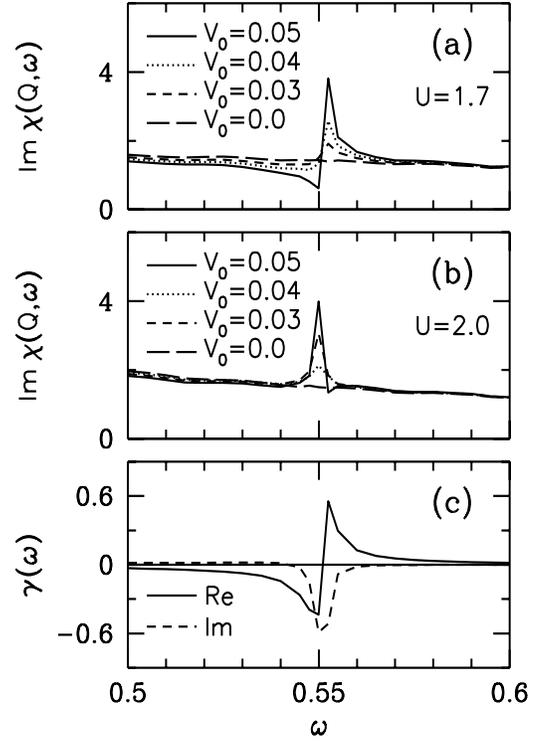}
\end{center}
\caption{
${\rm Im}\,\chi({\bf Q},\omega)$ versus $\omega$ for different 
values of $V_0$, and for 
(a) $U=1.7$ and (b) $U=2$.
(c) $\gamma(\omega)$ versus $\omega$ for $V_0=0.05$ and $U=2.0$.
These results were obtained for $\delta=0.001$.
}
\end{figure}
\noindent
by the spin fluctuations 
could have an effect similar to that of the 
finite broadening $\delta$ used here. 
This could be another reason for why the peak observed in 
${\rm Im}\,\chi({\bf Q},\omega)$ of Zn
substituted samples has a finite width. 
So, in these figures
it is observed that the quantitative features of 
the changes induced by the umklapp processes 
depend on the model parameters.

\section{DISCUSSION}

In summary, the effects of dilute nonmagnetic impurities on the 
frequency dependence of 
${\rm Im}\,\chi({\bf Q},\omega)$ have been studied.
These calculations were motivated by the neutron scattering data 
of Refs.~\cite{Sidis,Fong}
on 2\% and 0.5\% Zn substituted YBa$_2$Cu$_3$O$_7$.
The origin of the peak and of the low-frequency spectral 
weight induced by the impurities 
are the two issues which have been addressed in this paper.

In pure YBa$_2$Cu$_3$O$_7$, a resonant peak 
is observed only in the superconducting state at 
$\omega=41$ meV \cite{Rossat-Mignod,Mook,Fong95}.
The width of this peak is resolution limited 
as opposed to that observed in the 
Zn substituted samples. 
The theories attribute the resonant peak of the 
pure sample to instabilities in the particle-particle 
\cite{Demler,Demler2} 
or the magnetic channels \cite{Liu,Mazin,Bulut2}.
The calculations presented here 
\begin{figure} 
\begin{center}
\leavevmode
\epsfxsize=9cm 
\epsfysize=10cm 
\epsffile{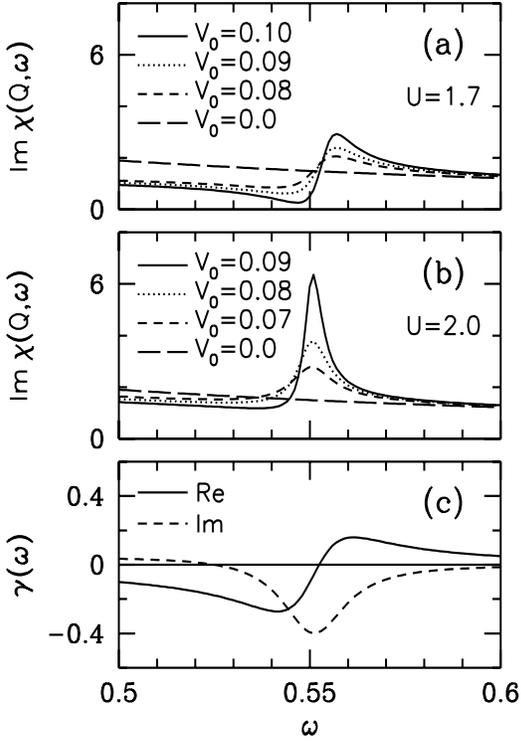}
\end{center}
\caption{
Results similar to those shown in Fig.~10 but for $\delta=0.01$
and different values of $V_0$.
}
\end{figure}
\noindent
show how a peak in 
${\rm Im}\,\chi({\bf Q},\omega)$ of Zn substituted YBa$_2$Cu$_3$O$_7$
could arise from the magnetic channel in the normal state.
However, clearly, the contributions to the peak from other
channels are not ruled out. 

The calculations have been carried out first without taking
into account the umklapp scattering of the spin fluctuations.
Here, the influence of the range of the impurity potential 
on $\chi_0({\bf Q},\omega)$ has been studied.
When a strongly attractive onsite impurity potential is used, 
it has been found that 2\% impurities have a negligible effect
on $\chi_0({\bf Q},\omega)$. 
This is in agreement with the calculations of 
Ref.~\cite{Li} in the unitary limit for an onsite potential. 
When an extended impurity potential is used with parameters 
similar to those obtained from the exact 
diagonalization calculations \cite{Ziegler}, 
2\% impurities lead to the smearing of 
$\chi_0({\bf Q},\omega)$ inducing spectral weight 
below $\omega_0$.
However, 
for 0.5\% impurities a negligible effect is found.

In the third section, 
the effects of the processes where the spin fluctuations scatter 
from the impurities with finite momentum transfers 
were taken into account.
It has been shown that the scatterings
of the spin fluctuations
with momentum transfers
${\bf Q^*}\approx 2{\bf k}_F$ lead to a peak in 
${\rm Im}\,\chi({\bf Q},\omega)$ at 
$\omega\approx\omega_0=2|\mu|$.
Here, the dependence of the impurity induced changes 
on the model parameters have been studied 
in order to make comparisons with the neutron scattering data. 
For instance, it was seen 
that the line shape of the peak depends 
on the model parameters.
When the parameters are such that the Stoner enhancement is small,
a dip is observed below the peak.
On the other hand, when 
$1-U{\rm Re}\,\chi_0({\bf Q},\omega)$ is small,
the dip is not observed and
the $2{\bf k}_F$ scatterings lead to a peak 
in ${\rm Im}\,\chi({\bf Q},\omega)$ by suppressing 
the damping of the RPA enhancement.
In addition, in this case, the width of the peak increases
with the broadening of the single-particle excitations.
It has been also found that the impurity scatterings 
with finite momentum transfers
could lead to spectral weight below $\omega_0$.
Hence, along with the results seen in the second section, 
the umklapp processes could also
play a role in inducing the 
low frequency spectral weight observed in 
2\% Zn substituted YBa$_2$Cu$_3$O$_7$ \cite{Sidis}.

The general features of the impurity induced changes in 
${\rm Im}\,\chi({\bf Q},\omega)$ calculated here appear 
to be in agreement with the experimental data on Zn substituted 
YBa$_2$Cu$_3$O$_7$.
However, in obtaining these results the Coulomb 
correlations were treated within RPA and 
$\chi_0({\bf Q},{\bf q},\omega)$ has been calculated 
in the lowest order in the strength of the impurity potential.
Furthermore, a static effective impurity potential was used. 
If the results presented here are supported 
by higher order calculations, then it would mean 
that a contribution to the peak observed in 
${\rm Im}\,\chi({\bf Q},\omega)$ could arise 
from the magnetic channel.
Furthermore, in this case, the experimental data 
of Refs.~\cite{Sidis,Fong}
would mean that a perturbation in the density channel 
as Eq.~(\ref{V2}) induces important changes 
in the antiferromagnetic response of YBa$_2$Cu$_3$O$_7$.

\acknowledgments

The author thanks P. Bourges, H.F. Fong and B. Keimer
for helpful discussions.
The numerical computations reported in this paper were performed 
at the Center for Information Technology at Ko\c{c} University.


    
\end{document}